\documentclass[rnote]{aa} 
\def\kt{\object{K~3-22}}
\def\ha{H$\alpha$}
\def\hb{H$\beta$}
\def\nii{[N\,{\sc ii}]}
\def\sii{[S\,{\sc ii}]}
\def\siii{[S\,{\sc iii}]}
\def\ariv{[Ar\,{\sc iv}]}
\def\ariii{[Ar\,{\sc iii}]}
\def\oiii{[O\,{\sc iii}]}
\def\oii{[O\,{\sc ii}]}
\def\oi{[O\,{\sc i}]}
\def\oip{O\,{\sc i}}
\def\ovip{O\,{\sc vi}}
\def\heii{He\,{\sc ii}}
\def\feii{[Fe\,{\sc ii}]}
\def\cav{[Ca\,{\sc v}]}
\def\fevii{[Fe\,{\sc vii}]}
\def\hei{He\,{\sc i}}
\def\hi{H\,{\sc i}}
\def\kms{\relax \ifmmode {\,\rm km\,s}^{-1}\else \,km\,s$^{-1}$\fi}
\usepackage{graphicx}
\usepackage{longtable,lscape}
\usepackage{txfonts}
\begin{document}
\title{\kt: a D-type symbiotic star \thanks{Based on observations
    obtained at the 2.5m~INT telescope of the Isaac Newton Group of
    Telescopes in the Spanish Observatorio del Roque de Los Muchachos
    of the Instituto de Astrof\'\i sica de Canarias.  This
      publication makes use of data products from the Two Micron All
      Sky Survey, which is a joint project of the University of
      Massachusetts and the Infrared Processing and Analysis
      Center/California Institute of Technology, funded by the
      National Aeronautics and Space Administration and the National
      Science Foundation.}}
\author{R.L.M. Corradi\inst{1,2}
           \and
        C. Giammanco\inst{1,2}
	}

   \offprints{R. Corradi}

   \institute{
Instituto de Astrof{\'{\i}}sica de Canarias, E-38200 La Laguna, 
Tenerife, Spain \email{rcorradi@iac.es}
   \and
Departamento de Astrof{\'{\i}}sica, Universidad de La Laguna, 
E-38205 La Laguna, Tenerife, Spain 
             }

\date{Received 6 August 2010 / Accepted 01 September 2010}

\abstract
{A goal of the IPHAS survey is to determine the frequency and nature
  of emission-line sources in the Galactic plane.}
{According to our selection criteria, \kt\ is a candidate symbiotic
star, but it was previously classified as a planetary nebula.}
{To determine its nature, we acquired a low-resolution optical spectrum
of \kt.}
{Our analysis of our spectroscopy demonstrates that \kt\ is indeed a
D--type symbiotic star, because of its high excitation nebular
spectrum and the simultaneous presence of Raman-scattered
\ovip\ emission at 6825~\AA\ and 7082~\AA, which is detected primarily
in symbiotic stars.}
{}
 \titlerunning{\kt: a D-type symbiotic
  star} \authorrunning{Corradi \& Giammanco} \maketitle


\section{Introduction}

While analysing the data of the the INT Photometric \ha\ Survey of the
northern Galactic plane (IPHAS, \cite{d05}), we noticed that a
point source included in the main list of Galactic planetary nebulae
(\cite{a92}), namely PN G045.6+01.5 (also called \kt), had optical and
near--infrared colours typical of symbiotic stars rather than
planetary nebulae (\cite{c08}, \cite{c10}).  We therefore decided to
investigate the nature of the source via optical spectroscopy. This
demonstrates that \kt\ is indeed a symbiotic star, as described in the
following.

\begin{figure*}[!ht]
\centering
\includegraphics[width=18.5cm]{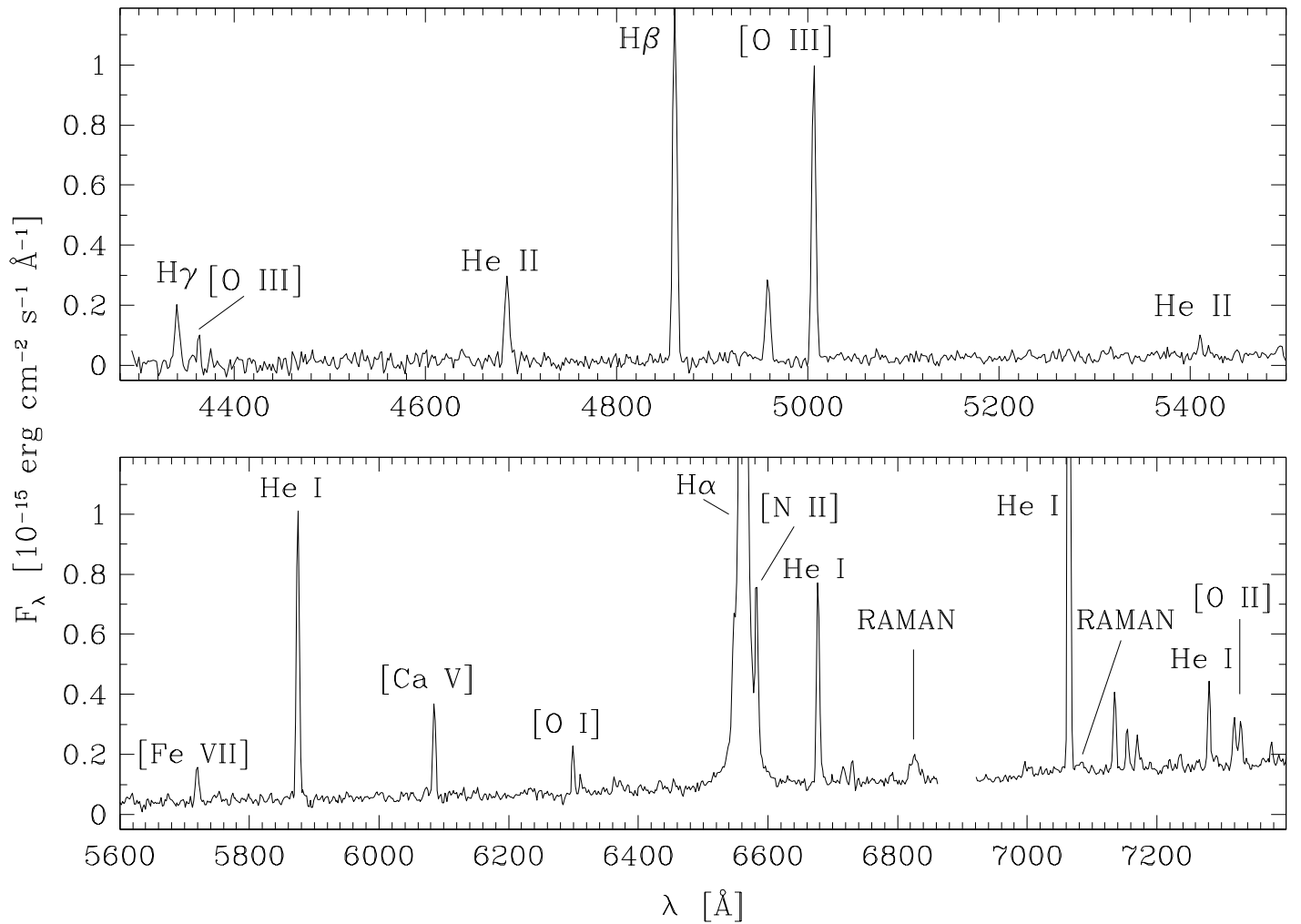}
\caption{The spectrum of \kt\ in the range with reliable flux
  calibration. The main spectral features are labelled. The
    region around 6870~\AA\ containing the telluric O$_2$ atmospheric
    absorption $B$ band is not plotted.}
\label{F-spectrum}
\end{figure*}

\section{Spectroscopic observations}

The spectrum of \kt\ was obtained on 2 September 2009 at the
2.5m~Isaac Newton Telescope using the IDS spectrograph.  Grating R300V
was used, which gives a reciprocal dispersion of 1.9~\AA\ per pixel,
and a spectral coverage from 4300 to 8500~\AA. The slit width was
1$''$.1, providing a spectral resolution of 5.0~\AA. The total
integration time on target was 80~min, which was divided into two
exposures of 40 min each.  The 2kx4k IDS EEV CCD is affected by
fringing redward of $\sim$7000~\AA. In addition, flux calibration is
uncertain above 7500~\AA\ because of second-order contamination as
well as significant optical aberrations at the edge of the
large-format CCD used with IDS.  Two spectrophotometric standards were
observed during the night to help us perform a relative flux
calibration. Reduction was completed using IRAF\footnote{IRAF is
  distributed by the NOAO which is operated by AURA under contract
  with NFS.} in a standard fashion.

\section{A symbiotic star's spectrum}

Figure~\ref{F-spectrum} shows the optical spectrum of \kt, which is
dominated by emission lines of both low (e.g., \oi, \feii, \sii) and
high (\oiii, \heii, \cav, \fevii) ionization species. \hi\ and
\hei\ emission is strong, and \heii\ relatively fainter. Line
identification and fluxes are listed in Table~\ref{T-emlines}. Line
fluxes F$_{obs}$ are given relative to F$_{H\beta}$=100, and the
estimated \hb\ flux is 6.5$\times$$10^{-15}$~erg~cm$^{-2}$~s$^{-1}$,
with an error of 20\%\ caused by a lack of precise absolute flux
calibration for these data. The errors in the quoted relative fluxes
are $\sim$5\%\ for lines stronger than 0.2\,F$_{H\beta}$, and are
larger for fainter lines. The \ha\ and in particular the
\nii6583 fluxes are affected by an additional amount of uncertainty 
because of their blending.

Relatively strong \oip\ emission at 8446~\AA\ is also detected, but
this line is outside the spectral range with reliable flux
calibration. The continuum is weak and slowly rising, with no evidence
of the absorption bands of a red giant. However, the clear detection
of the Raman--scattered \ovip\ broad emission at 6825~\AA\ and the
fainter line 7082~\AA\ (\cite{sm01}), in addition to high excitation
lines, proves its nature as a symbiotic star according to the criteria
defined in \cite{b00}.  The full width at half maximum (FWHM) of the
Raman-scattered lines is 15~\AA.  The classification of \kt\ as a
symbiotic star is also supported by the large observed \ha/\hb\ flux
ratio, of the order of 50, which is typical of symbiotic binaries.
The broad wings of the \ha\ line profile of \kt, which extend well
beyond the \nii\ doublet (see Fig.~\ref{F-spectrum}) with a total
width at zero intensity of 4500~\kms, are also characteristic of
symbiotic stars (\cite{vw93}), and might also have been produced by
Raman scattering (\cite{n89}).

The situation is similar to that of other symbiotic systems, such as
He~2-104, where the red giant is not visible in the optical, and only
shows up in the near-IR either spectroscopically via its
characteristic CO absorption bands in the K band, or by photometric
variations typical of pulsating AGB stars (\cite{sg08}).  These
systems generally belong to the class of ``dusty'' (D--type) symbiotic
stars, because their near-IR colours are indicative of a reddened
Mira-type giant surrounded by a large amount of warm (T$\sim$1000~K)
dust. This is also true for \kt: its 2MASS magnitudes are
$J=12.34$, $H=10.33$, and $K=8.82$~mag, which mean that the source is
at the top-right of the locus defined by D-type symbiotic stars in the
2MASS colour-colour diagram (\cite{c08}).

Therefore these observations resolve the previous uncertainty about
the nature of \kt: the source was included in the list of true or
probable planetary nebulae (PNe) in the Strasbourg--ESO catalogue but
with the note ``possibly an H\,{\sc ii} region'' (\cite{a92}),
classified as a possible planetary nebula (\cite{k01}), or considered
as an object of uncertain nature (\cite{a87}).

To summarise, \kt\ is one of a group of elusive D-type symbiotic stars
for which the red giant in the optical range is only indirectly
detected by the excitation of the circumstellar nebula, which allows
the formation of Raman-scattered emission thanks to the simultaneous
presence of high energy photons from the hot white dwarf companion and
neutral material in the red-giant extended atmosphere.  The
observational frontier between these systems and a number of PNe is
poorly defined; in both cases, the optical spectrum is dominated by
the ionized nebula, and the central star(s) are highly obscured by
dust. Both classes also often display extended ionized nebulae with
similar bipolar morphologies (\cite{c03}).  The near-IR can provide
clues about their symbiotic nature, but not without difficulties
(cf. \cite{sg09}).  Even the presence of the Raman-scattered
\ovip\ broad emission at 6825~\AA\ is not a property exclusive to
symbiotic stars: it has also been detected in bipolar PNe with massive
neutral envelopes such as NGC 6302 (\cite{gr02}) and NGC 7027
(\cite{z05}), although in the latter objects they are much fainter
than the nebular emission lines. Unveiling the true nature of
these and other well-studied nebulae such as M~2--9 or Mz~3 (see
\cite{c95}), or of certain newly discovered candidate PNe
(e.g. \cite{v09}), remains a difficult task, but is of vital
importance to understanding the physical processes at work in forming and
shaping the outflows from evolved stars.

\begin{acknowledgements}
This work is supported by the Spanish AYA2007-66804 grant.
\end{acknowledgements}

\begin{table}
  \caption{Emission line identification and fluxes, relative
    to F$_{H\beta}$=100.}
\begin{tabular}{llr}
\hline\hline
  Identification & $\lambda [\AA]$       & F$_{obs}$ \\
\hline\\[-4pt]
\hi\         & 4340.5 &    18 \\ 
\oiii\       & 4363.2 & $^\star$ \\ 
\heii\       & 4685.7 &    26 \\ 
\hi\         & 4861.4 &   100 \\ 
\oiii\       & 4958.9 &    24 \\ 
\oiii\       & 5006.8 &    81 \\ 
\heii\       & 5411.5 &     5 \\ 
\fevii\      & 5720.9 &    11 \\ 
\hei\        & 5875.6 &    80 \\ 
\cav/\fevii\ & 6086   &    26 \\ 
\oi\         & 6300.3 &    11 \\ 
\siii\       & 6312.1 &     4  \\ 
\hi\         & 6562.8 &  5500$:$ \\ 
\nii         & 6583.4 &   60$:$ \\ 
\hei\        & 6678.2 &    55 \\ 
\sii\        & 6716.4 &     5 \\ 
\sii\        & 6730.8 &     4 \\ 
Raman        & 6825   &    19 \\ 
\hei\        & 7065   &   218 \\ 
Raman        & 7082   &     6 \\ 
\ariii\      & 7135.8 &    20 \\ 
\feii\       & 7155.1 &    11 \\ 
\feii\       & 7172.0 &     9 \\ 
\ariv\       & 7236.0 &     4 \\ 
\hei\        & 7281.4 &    21 \\ 
\oii\        & 7319   &    16 \\ 
\oii\        & 7330   &    13 \\[2pt]
\hline\\[-4pt]                             
\end{tabular}
\newline $^\star$ marginal detection\\
$:$ uncertain flux because of blending.
\label{T-emlines}
\end{table}

\end{document}